\documentclass[12pt]{iopart}

\usepackage{iopams}  
\expandafter\let\csname equation*\endcsname\relax
\expandafter\let\csname endequation*\endcsname\relax 
\usepackage{amsmath,amssymb}
\usepackage{bbold}
\usepackage{upgreek}
\usepackage{dsfont}
\usepackage{graphicx, epstopdf} 
\usepackage{amsfonts}
\usepackage{graphicx,xcolor}
\usepackage{amsmath,bbm}
\usepackage{braket}
\usepackage[utf8x]{inputenc}
\usepackage[ngerman, english]{babel}
\usepackage[T1]{fontenc}
\usepackage{mathtools}
\usepackage{hyperref}
\usepackage{soul} 
\usepackage{cite}

\begin{document}


\title{A Comprehensive Study on A Tapered Paul Trap: From Design to Potential Applications}

\author{Bo Deng$^{1}$, Moritz G{\"o}b$^1$, Max Masuhr$^{1,2}$, Johannes Ro{\ss}nagel$^3$, Georg Jacob$^4$, Daqing Wang$^{1,2,\dag}$ and Kilian Singer$^{1,\ddag}$}

\address{$^1$ Experimental Physics I, Institute of Physics, University of Kassel, Heinrich-Plett-Stra{\ss}e 40, 34132 Kassel, Germany}
\address{$^2$ Institute of Applied Physics, University of Bonn, Wegelerstra{\ss}e 8, 53115 Bonn, Germany}
\address{$^3$ Holunderweg 4, 55299 Nackenheim, Germany}
\address{$^4$ Innstraße 109, 6020 Innsbruck, Austria}
\ead{$^\dag\,$daqing.wang@uni-bonn.de, $^\ddag\,$ks@uni-kassel.de}

\begin{abstract}
We present a tapered Paul trap whose radio frequency electrodes are inclined to the symmetric axis of the endcap electrodes, resulting in a funnel-shaped trapping potential. With this configuration, a charged particle confined in this trap has its radial degrees of freedom coupled to that of the axial direction. The same design was successfully used to experimentally realize a single-atom heat engine, and with this setup amplification of zeptonewton forces was implemented. In this paper, we show the design, implementation, and characterization of such an ion trap in detail. This system offers a high level of control over the ion's motion. Its novel features promise applications in the field of quantum thermodynamics, quantum sensing, and quantum information.
\end{abstract}
\maketitle

\section{Introduction}
Trapped ions possess unrivaled features covering a broad range of fields such as quantum metrology and sensing \cite{Ludlow2015, Hempel2013, Wolf2021, Milne2021}, quantum information \cite{Monroe2013,Mehta2020-ur,Jain2024-sd}, quantum simulation \cite{Monroe2021} and quantum thermodynamics \cite{Pijn2022, Kranzl2023}, to name a few. The geometry of ion traps varies from macroscopic three-dimensional linear Paul traps \cite{Blatt2012}, needle traps \cite{wang2016fabrication}, segmented micro-traps \cite{decaroli2021design,bautista2019}, to surface traps \cite{lekitsch2017blueprint, wright2013reliable} tailored to different applications. Here, we present and characterize a macroscopic tapered Paul trap that possesses distinct properties that have been exploited for studying nonlinear mechanical oscillators with applications for force sensing and amplification \cite{Deng2023}. The special features due to the tapered design have also been used to implement a single atom heat engine \cite{Rossnagel2016} with the potential to implement a thermal machine in the quantum regime \cite{Bouton2021}.\\
This article begins with an introduction to the design and construction of this tapered ion trap. We then proceed to a numerical field simulation and a 3D trajectory simulation. These simulations allow us to evaluate the performance of our trap. Following this, we provide a detailed description of the experimental setup and present characterizations of this system. Our measurements characterize the axial position dependent radial confinement as a special feature of this system. We also present sideband spectroscopy performed on the $4^2S_{1/2} \leftrightarrow 3^2D_{5/2}$ quadrupole transition of $^{40}$Ca$^+$ ions using the electron shelving technique. Finally, we discuss the potential applications of this system in various fields, including quantum thermodynamics, quantum metrology and sensing, and quantum information.\\

\section{Trap design and field simulation}
\label{sec:Sim}

\begin{figure}[h]
\begin{center}
\includegraphics[width=0.8\columnwidth]{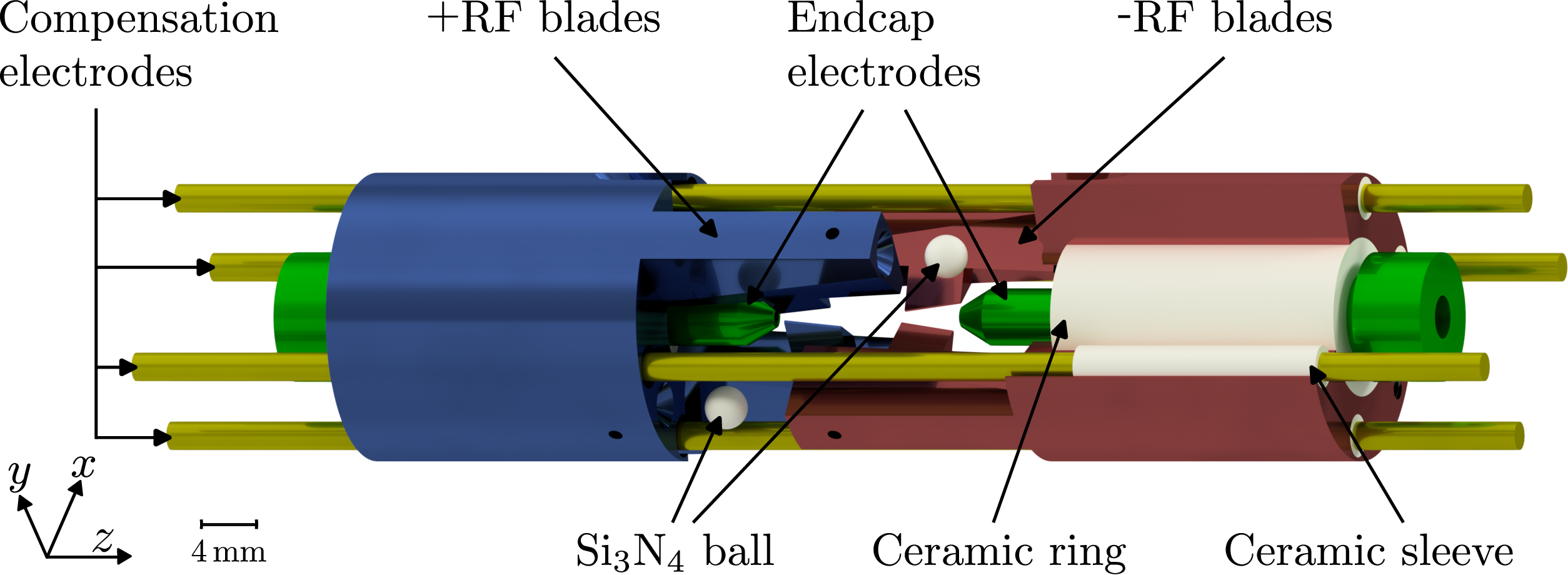}
\caption{\label{fig:explodedcad}An exploded view of the tapered Paul trap design. The fork-shaped RF blade electrodes (red and blue) are inserted into each other at the orthogonal orientation in $x-y$ plane. The Si$_3$N$_4$ balls between the RF electrodes are for precise positioning during the trap assembly as well as electric insulation. Endcap electrodes (green) are inserted along the symmetry axis and insulated with ceramic rings. Four compensation electrodes (yellow) are located in the outer vicinity of the assembly stack.}
\end{center}
\end{figure}
Conventional linear Paul traps, as used in many state-of-the-art quantum information experiments \cite{Blatt2012, Wineland2013, Monroe2021}, are designed and constructed to minimize couplings between motional degrees of freedom. However, such couplings can be beneficial in several research areas, such as studies with single-atom heat engines \cite{Abah2012, Rossnagel2016} in the quantum regime, quantum metrology experiments, and exploring new methods for quantum information. To facilitate this coupling, we deviate from the conventional design of the Paul trap by inclining the four blade-shaped radiofrequency (RF) electrodes to the axial symmetry axis by an angle $\vartheta=10\,^\circ$, which is chosen such that adequate radial-axial coupling is achieved while maintaining stable trapping conditions. Special care is taken to avoid axial micromotion by driving the four blades with RF voltages of opposite sign between each pair of blades (see Fig.\,\ref{fig:explodedcad}). Furthermore, micromotion is minimized by a proper alignment of the RF electrodes with respect to the axial $z$ axis of the trap, as shown in Fig.\,\ref{fig:explodedcad}. This is achieved by using four silicon nitride insulating balls and corresponding recesses to position the fork-shaped radio frequency electrodes precisely. 
The endcap electrodes are inserted into the RF electrodes and insulated from the latter by ceramic rings. Additionally, four compensation electrodes are inserted and insulated by ceramic sleeves. \\

\begin{figure}[h]
\begin{center}
\includegraphics[width=0.8\linewidth]{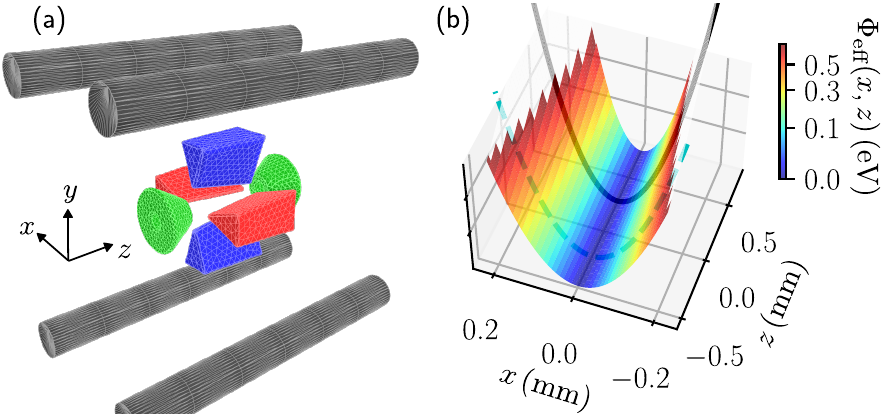}
\caption{\label{fig:cad}Pseudopotential simulations: (a) Simplified trap geometry for numerical simulations. The RF blades (blue and red) with an axial length of 4\,mm are driven with a voltage of opposite sign. The endcap electrodes (green) spaced by 4.8\,mm provide confinement along the $z$-direction by applying DC voltages. The compensation electrodes (grey) are responsible for micromotion compensation (diagonal distance 17\,mm and 2\,mm diameter). (b) The pseudopotential in the $x$-$z$-plane. The black solid and cyan dashed lines illustrate a fitted harmonic potential, showing the axial position dependent confinement.
}
\end{center}
\end{figure}

The RF blades have a length of 4$\,$mm, and the endcap electrodes are spaced 4.8$\,$mm apart. This yields a factor of 16 smaller trapping volume compared to the tapered Paul trap used in our earlier work\,\cite{Rossnagel2016}. The endcap electrodes are narrowed towards the center of the trap and have a through-hole of 0.8$\,$mm diameter, allowing for optical access along the axial axis. The electric fields generated by this geometry are simulated using a boundary element method\,\cite{Singer2010, Betcke2021}. In order to solve the potential of the trap, the computer-aided design files, including the blade-electrode pairs, the endcap electrodes, and the compensation electrodes, are imported. The polygon mesh in Fig.\,\ref{fig:cad} (a) used for this numerical simulation is generated by the finite element mesh generator Gmsh \cite{gmsh}. Based on the imported geometry, the electrostatic potential is solved using the boundary element solver package Bempp \cite{Betcke2021}. By applying the pseudopotential approximation \cite{Leibfried2003, Singer2010}, the trapping potential can be described by $\Phi_\text{eff}(x,y,z)=q\left|\nabla \Phi(x,y,z)\right|^2/\left(4m\omega_\text{RF} ^2\right)$ \cite{InhomogenousRFFields1992} with $\Phi(x,y,z)$ the electrostatic potential, $q$ the charge of the particle, $m$ its mass and $\omega_\text{RF}$ the RF frequency. In Fig.\,\ref{fig:cad}(b), a cross-section of the pseudopotential in the $x$-$z$-plane is depicted. The confinement in the radial direction is illustrated by the black solid and cyan dashed lines, which are obtained by fitting a quadratic function to the potential. The confinement is getting stronger towards positive $z$. To understand the behavior of a single ion inside the potential, we perform a trajectory simulation of a single calcium ion using the \textit{velocity Verlet} propagator \cite{Singer2010}. We set the voltages on the two pairs of the blade electrodes as $\Tilde{V}_\text{RF1}=\hat{V}_\text{RF1}\cos(\omega_\text{RF}t)$ and $\Tilde{V}_\text{RF2}=-\hat{V}_\text{RF2}\cos(\omega_\text{RF}t)$ respectively, where $\hat{V}_\text{RF1,2}$ denotes the amplitude of the RF voltages. We deviate $\hat{V}_\text{RF1}$ and $\hat{V}_\text{RF2}$ by 1.8\,\% to mimic the imperfect symmetry of the experimental setup, which lifts the degeneracy of the two radial modes. By varying the voltages on the endcaps, we can shift the potential minimum along the axial direction. We repeat the simulation with different axial potential minimum positions and perform a Fourier transformation to the simulated trajectories in $x$ and $y$ to determine the corresponding trapping frequencies. The simulation result is compared to measurements in Sec.\,\ref{sec:Char}.\\

\section{Experimental setup}

\begin{figure}[h]
\begin{center}
\includegraphics[width=0.7\columnwidth]{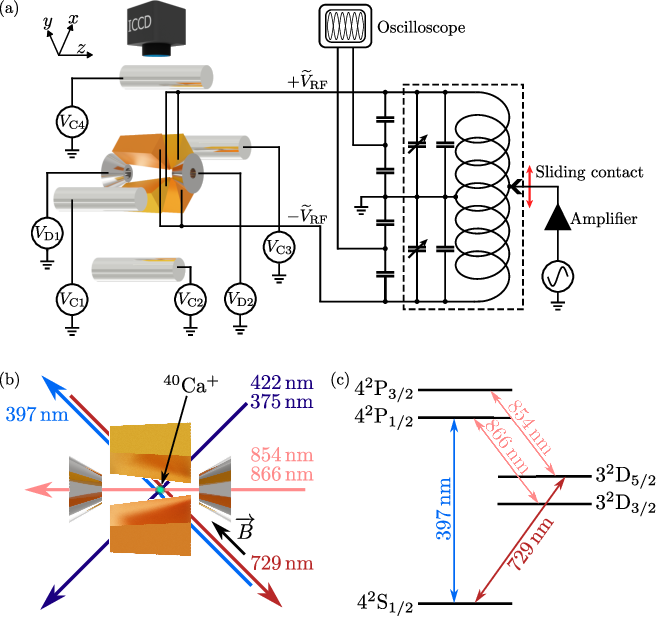}
\caption{Experimental setup of the tapered Paul trap. (a) A helical resonator provides a symmetric radio frequency drive to the RF blades avoiding excessive micromotion in the axial direction. DC-voltages $V_\textrm{D1}$ and $V_\textrm{D2}$ are applied to end-caps. The voltages applied to the compensation electrodes ($V_\textrm{C1}$ to $V_\textrm{C4}$) are used for micromotion compensation and radial principal axes adjustment. An intensified CCD camera is used to image single ions with a viewing direction perpendicular to the $z$ axis and at 45 degrees with respect to the $x$ and $y$ axis. (b) Top view on the trap: Calcium atoms are provided by a thermal source. The laser system consists of ionization lasers (at 422$\,$nm and 375$\,$nm) with cooling (at 397$\,$nm) and repumping lasers (at 854$\,$nm and 866$\,$nm) for Doppler cooling. A quantization axis is defined by applying 3$\,$G magnetic field by a pair of permanent magnet rings counter-propagating with a narrow band 729$\,$nm laser used for the $4^2S_{1/2}$ to $3^2D_{5/2}$ quadrupole transition. (c) Relevant energy levels and optical transitions of the $^{40}$Ca$^{+}$ ion.}
\label{fig_experiment_setup}
\end{center}
\end{figure}
Figure\,\ref{fig_experiment_setup}\,(a) shows the electric connections to the Paul trap. The dynamic confinement in the radial directions ($x$, $y$) is provided by RF voltages ($+\widetilde{V}_{\mathrm{RF}}$, $-\widetilde{V}_{\mathrm{RF}}$) applied to the two opposite-positioned pairs of blades, respectively. An RF signal at 11.17$\,$MHz from an analog
signal generator is amplified by a 5-watt power amplifier with a gain of 46\,dB. The amplified signal is then sent to a helical resonator for further amplification and impedance matching to the trap. It is specifically designed with a $\lambda/2$ geometry\,\cite{RoßnagelDiss}, whose middle point is grounded with two open ends connected to the two pairs of RF blades. This configuration provides two RF voltages with a $\pi$ phase difference that is necessary to avoid excessive axial micromotion. The RF voltages are probed by voltage dividers composed of two capacitors with a ratio of 118 for ease of characterization and adjustment. Through careful adjustment of the sliding contact on the coil and the tunable capacitors on the helical resonator, we tune to an impedance-matched frequency of 11.17$\,$MHz, where the quality factor of this helical resonator together with the Paul trap is measured by a vector network analyzer to be about 46. Through analyzing the probed RF signal on the oscilloscope, the phase difference of the two RF signals is measured to be 179.51($\pm0.28$) degrees with a relative amplitude difference of 1.29\%. The amplitudes of both probed RF signals are around 0.8\,V. The RF voltages on the blade-shaped RF electrodes have amplitudes of around 95\,V. The axial confinement is achieved by applying DC voltages ($V_{\mathrm{D1}}$ and $V_{\mathrm{D2}}$) to the endcap electrodes. Additionally, four auxiliary electrodes are implemented around the Paul trap (see Fig.\,\ref{fig:explodedcad} in yellow). These electrodes are supplied with four independent DC voltage sources, which allow for rotating radial principal axes and compensation of the micromotion\,\cite{berkeland1998minimization}.\\
The trap is contained in an ultrahigh-vacuum chamber. In our experiment, $^{40}\text{Ca}^{+}$ ions are used. The calcium sample is sublimated in an oven to form a collimated atomic beam shooting through the center of the trap. The atoms are ionized via a two-photon process by the combination of a 422$\,$nm and a 375$\,$nm laser\,\cite{gulde2001simple}. A 397$\,$nm laser red-detuned from the $4^2S_{1/2} \leftrightarrow 4^2P_{1/2}$ transition of $^{40}\text{Ca}^{+}$ is focused into the trap for Doppler cooling. The short lifetime of the $4^2P_{1/2}$ state leads to a natural linewidth of 22.1$\,$MHz. As the ion at state $4^2P_{1/2}$ has a probability of 0.064 to decay to the metastable state $3^2D_{3/2}$ \cite{ramm2013precision}, an 866$\,$nm laser is used to repump the ion back to the $4^2P_{1/2}$ state to close the laser cooling cycle.  A second repumper laser at 854$\,$nm is used for pumping the $3^2D_{5/2} \leftrightarrow 4^2P_{3/2}$ transition. The resonantly scattered photons at 397$\,$nm are used for fluorescence imaging. In addition, a 729$\,$nm laser is locked to a high finesse cavity. It is used to address the $4^2S_{1/2} \leftrightarrow 3^2D_{5/2}$ quadrupole transition to implement resolved sideband cooling. To define a quantization axis, permanent magnets are mounted outside the vacuum chamber to generate a magnetic field as shown in Fig.\,\ref{fig_experiment_setup}(b), which also lifts the degeneracy of magnetic quantum states.\\

\section{System characterization}
\label{sec:Char}
\begin{figure}[h]
\begin{center}
\includegraphics[width=1.0\columnwidth]{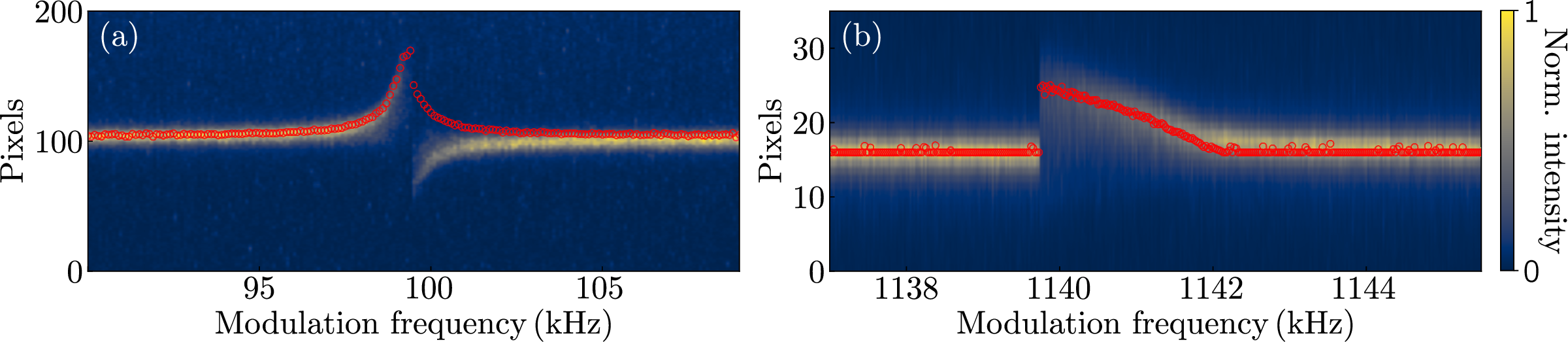}
\caption{\label{fig:epsart} (a) Motional mode measurement of a single ion excited by intensity modulation of the 397\,nm laser for frequencies around the axial resonance. The red circles show the oscillation amplitude at each frequency.
(b) Similar procedure with intensity modulation of the 397\,nm laser around the radial trapping frequency showing a Duffing-type nonlinear response.}
\label{axial_mode_meas}
\end{center}
\end{figure}

\begin{figure}[h]
\centering
\includegraphics[width=0.5\columnwidth]{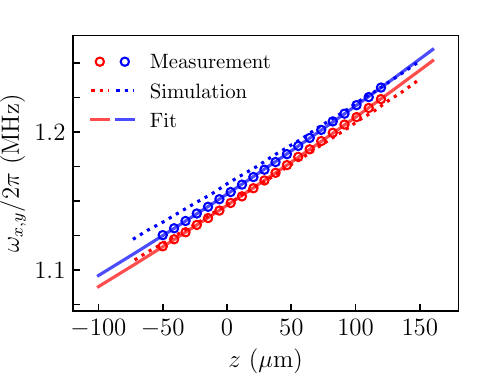}
\caption{Axial position dependent radial confinement. The red (blue) circles represent the measured trapping frequencies in the $x$-($y$-)direction at different axial positions fitted to Eq.\,(\ref{analytical_axial_dependent_radial}) (solid lines). The dotted lines show the result of the trajectory simulation in the respective direction as described in Sec.\,\ref{sec:Sim}. }
\label{fig:radFreq_vs_zpos}
\end{figure}

Next, we measure the trapping frequency of a single $^{40}\text{Ca}^{+}$ ion by exciting its motion with a periodic force. To do so, we red-detune the 397$\,$nm laser by 33.7$\,$MHz from the $4^2S_{1/2}\leftrightarrow4^2P_{1/2}$ transition and apply a sinusoidal modulation to its intensity. We monitor the fluorescence from the ion as a function of the modulation frequency. Resonant scattering of photons excites the mechanical oscillation when the intensity modulation frequency matches one of the harmonic trapping frequencies. Figure\,\ref{axial_mode_meas}(a) displays the fluorescence signal from the ion as a function of intensity modulation frequency during an axial trapping frequency measurement. Here, the fluorescence signal is imaged by a camera with an exposure time of 400$\,$ms at each frequency. Each column of the displayed image represents a summation of a fluorescence image over the radial directions. As the intensity modulation frequency is tuned close to the axial trapping frequency at 99.8\,kHz, a coherent oscillation along the axial direction is excited. Since the exposure time of the camera is much longer than one oscillation period, the recorded images integrate the fluorescence signal over many oscillation cycles, which can be formulated as the position distribution function of sinusoidal oscillation convoluted with the Gaussian point spread function of the imaging system. The oscillation amplitude (red circles) is obtained by fitting each column to this model. Figure\,\ref{axial_mode_meas}(b) shows the fluorescence signal from the ion when one of the radial oscillation modes is excited. Here, each column represents an integration of the image over the axial direction. The oscillation amplitude (red circles) in Fig.\,\ref{axial_mode_meas}(b) is obtained following the same procedure as in Fig.\,\ref{axial_mode_meas}(a).\\
The red and blue circles in Fig.\,\ref{fig:radFreq_vs_zpos} show the measured trapping frequencies of the $x$ and $y$ radial modes as a function of the axial position of the ion. To obtain the data, we shift the axial position of the trapped $^{40}\text{Ca}^{+}$ ion by changing the voltages applied to the endcap electrodes ($V_{\mathrm{D1}}$,$V_{\mathrm{D2}}$), and keeping the axial trapping frequency constant. The radial frequency measurement is performed at each axial position by a descending sweep of the modulation frequency across the radial modes. The red and blue dotted lines display the results from trajectory simulations mentioned in Sec.\,\ref{sec:Sim}, showing good agreement with the measured data. The solid lines are from data fitting to the following analytical formula 

\begin{equation}
\label{analytical_axial_dependent_radial}
    \omega_{\ell}(z)=\dfrac{\omega_{\ell,0}}{\left(1+{z\tan{\vartheta}}/{r_0} \right)^2},
\end{equation}
where $\omega_{\ell}(z)$ are the two radial trapping frequencies with $\ell=\{x,y\}$ and $\omega_{\ell,0}$ denotes their value at $z=0$ and $r_0$ denotes the distance of the ion trap center to the center of RF blades.\\
In the experimentally measured range $z=[-50, 100]\,\mu m$, the dependence of the radial trapping frequency can be well approximated to be linear to the axial position, which can be described by $\omega_{l}=\omega_{l,0}{\left(1+\epsilon z\right)}$, with $l=\{x,y\}$, $\omega_{l,0}\simeq2\pi\times\{1.14,1.15\}$\,MHz and $\epsilon=0.552\,\text{mm}^{-1}$.\\

\begin{figure}[h]
\begin{center}
\includegraphics[width=1.0\columnwidth]{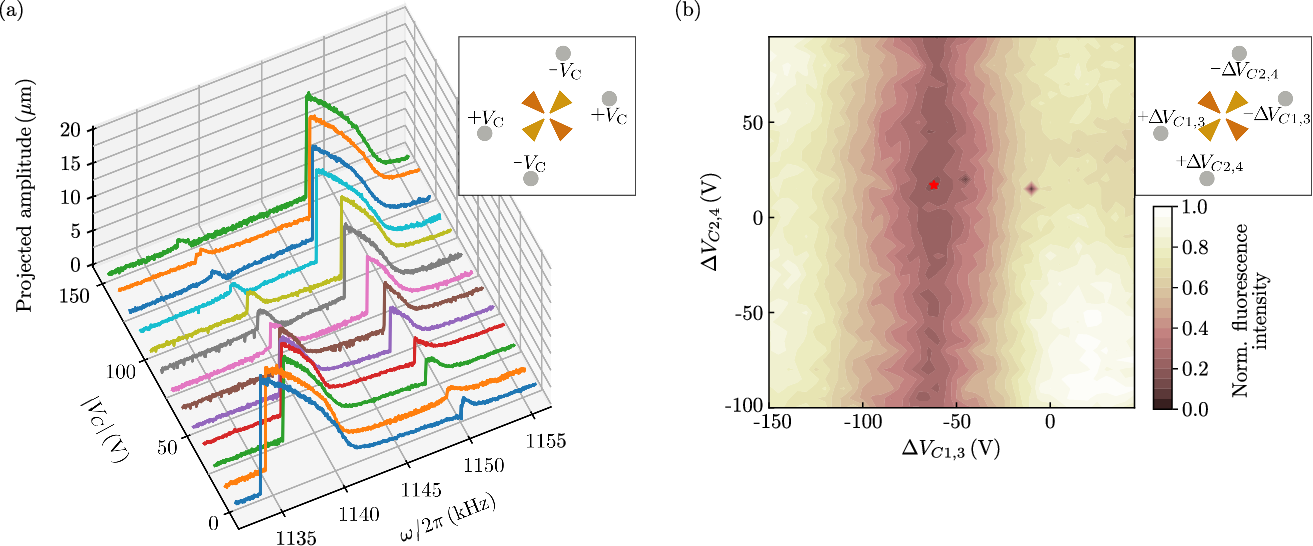}
\caption{\label{fig:epsart} (a) The oscillation amplitudes of the radial modes projected to the imaging plane are shown as a function of the voltage $V_\textrm{C}$, which is applied to the compensation electrodes as illustrated by the inset.  An increase in the voltage leads to a clockwise rotation of the principal axis when looking along the $z$-axis. We choose $V_\textrm{C}$ such that the amplitudes of the two radial modes are equal, as seen in the brown trace. (b) Based on the voltage settings acquired in (a), the micromotion compensation is performed by monitoring the fluorescence of the ion as a function of $\Delta V_\textrm{C2,4}$ and $\Delta V_\textrm{C1,3}$, which move the ion in the radial plane. Minimal values are obtained near the red star mark.}
\label{fig_prin_axis_rot}
\end{center}
\end{figure}

To obtain the amplitude of the coherent oscillation in the radial directions, we need to determine the angle between the imaging plane and the two radial principal axes. As shown in Fig.\,\ref{fig:radFreq_vs_zpos}, these two radial modes are non-degenerate by 8\,kHz. The orientation of principal axes can be controlled by voltages on the compensation electrodes (see Fig.\,\ref{fig:explodedcad}) to compress or stretch the trapping potential. The tapered geometry leads to an asymmetric shielding effect to the compensation electrodes. In order to maintain the axial position of the ion, the voltages of the four compensation electrodes are applied in a manner as shown in the inset of Fig.\,\ref{fig_prin_axis_rot}(a). We show the radial frequency response projected to the imaging plane at different values of the compensation voltage $V_C$ in Fig.\,\ref{fig_prin_axis_rot}(a). By increasing $V_C$ from 0 to 150 Volt, we can observe the radial principal axes rotating clockwise when looking along the $z$-axis.
The two radial principal axes both have an angle to the imaging plane of 45$\,^{\circ}$ when setting $V_C=62.5\,V$.\\

After the principal axes are set, micromotions of the ion can be compensated by shifting the ion's position in the radial plane by applying voltages on the compensation electrodes. To keep the axial position of the trapped ion, the changes in the voltages applied to the opposite-positioned electrodes need to be of opposite sign, as shown in the inset of Fig.\,\ref{fig_prin_axis_rot}(b). As the frequency of the micromotion is 11.17\,MHz, which is smaller than the natural linewidth of the $4^2S_{1/2}\leftrightarrow4^2P_{1/2}$ transition, the spectrum of this transition will be broadened due to excess micromotions \cite{berkeland1998minimization}. In this measurement, the laser is red detuned by 33\,MHz from the resonance. The micromotion is reduced by minimizing the fluorescence signal of the $4^2S_{1/2}\leftrightarrow4^2P_{1/2}$ transition\,\cite{berkeland1998minimization}. As shown in Fig.\,\ref{fig_prin_axis_rot}(b), the fluorescence signal is plotted against the voltages $\Delta V_{C1,3}$ and $\Delta V_{C2,4}$, which are set to 18.5\,V and -60\,V to minimize the micromotion.\\

\section{Sideband spectroscopy}
We have also measured the sideband-resolved spectrum of the $4^2S_{1/2} \leftrightarrow 3^2D_{5/2}$ quadrupole transition. The long lifetime of 1.05\,s \cite{james1998quantum} of the $3^2D_{5/2}$ state leads to a narrow bandwidth. A frequency-stabilized laser at 729\,nm with a nominal linewidth of 1\,Hz is used to address this transition. At the trap, 1.3$\,$mW of 729$\,$nm laser power is available, with a focal diameter at the ion of around 50$\,\mu$m. The laser frequency can be swept across a range of 20\,MHz with a double-pass acousto-optic modulator (AOM) setup, which is driven at a center frequency of 330$\,$MHz. The driving signal of the AOM is controlled by a fast RF switch. There are in total ten transitions between the different sublevels of $4^2S_{1/2}$ and $3^2D_{5/2}$, spanning 20$\,$MHz under the 3-Gauss magnetic field inside the trap. The 729\,nm laser orientation is chosen parallel to the quantization axis defined by the magnetic field [see Fig.\,\ref{fig_experiment_setup}(b)], selecting transitions with $|\Delta m_J|\,=\,$1 \cite{Roos}.\\
\begin{figure}[h]
\begin{center}
\includegraphics[width=0.6\columnwidth]{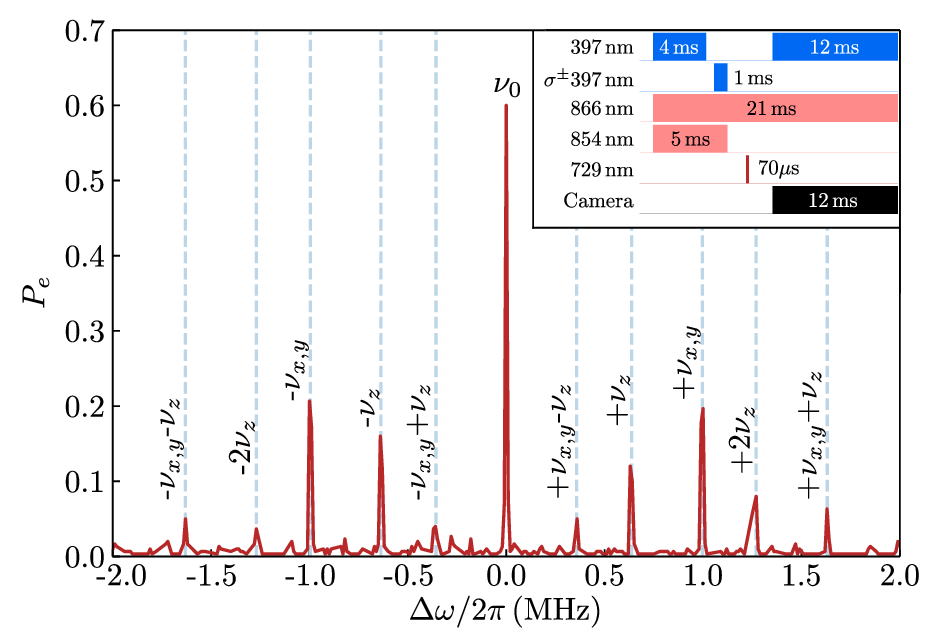}
\caption{Sideband excitation spectrum for the $|4^2S_{1/2},+1/2\rangle\leftrightarrow|3^2D_{5/2},-1/2\rangle$ transition. Each data point corresponds to 300 repetitions of the pulse sequence. The carrier transition is positioned at zero detuning. Red and blue sidebands for an axis $i$ are depicted with $\pm v_i$, respectively. The two radial axes $x,y$ are not distinguished in the spectrum. Hybrid coupled motional sidebands are also visible. Inset: Pulse sequence for obtaining the sideband spectrum. The ion is cooled with a linear 397 nm beam and initialized in the corresponding $m_J$-substate by a circularly polarized $\sigma^\pm$  397\,nm beam. A 729\,nm beam with a chosen detuning then drives the $4^2S_{1/2} \leftrightarrow 3^2D_{5/2}$ transition. The ion fluorescence is then measured with the camera under 397\,nm illumination.}
\label{fig_sidebandAndPulseSequence}
\end{center}
\end{figure}
The quadrupole excitation spectrum shown in Fig.\,\ref{fig_sidebandAndPulseSequence} is obtained by the electron shelving method\,\cite{bergquist1986observation}. The inset shows the pulse sequence of the measurement. The ion is initially laser-cooled using a linearly polarized 397\,nm laser pulse. It is then pumped by a circularly polarized 397$\,$nm pulse to the $|4^2S_{1/2},+1/2\rangle$ Zeeman state. Afterwards, a 729$\,$nm laser pulse is sent to the ion for 70$\,\mu$s. Finally, a 397$\,$nm pulse is turned on along with the camera to measure the fluorescence of the ion. The ion appears dark when the 729$\,$nm pulse excites the ion to the $D$-state. Otherwise, the ion appears bright on the camera. The spectrum of the $|4^2S_{1/2},+1/2\rangle\leftrightarrow|3^2D_{5/2},-1/2\rangle$ transition is obtained by sweeping the driving frequency of the AOM. At each frequency, the measurement is repeated 300 times. The 866$\,$nm laser beam is continuously on during the measurement. The 854$\,$nm laser beam is only used before the 729\,nm laser manipulation to initialize the ion to its $4^2S_{1/2}$ state. The strongest peak in the center of the spectrum is the carrier transition. To confirm that the remaining peaks are motional sidebands, the spectrum is taken with different axial trapping frequencies. Higher-order sidebands are also visible in the spectrum, as our system is not operating in the Lamb-Dicke regime. There are also hybrid motional sidebands where both the radial and axial motional modes are accessed.\\

\section{Discussion and Outlook}
A detailed description of the setup presented in this article allows analytical and numerical investigations of the trap and provides a sound basis for designing, optimizing, and tailoring properties of the system for new applications and basic research. We want to draw attention to our tapered Paul trap platform and spark interest to utilize its special features, such as mechanical nonlinearities, to implement novel schemes for quantum information, to advance metrology and sensing methods, and to push forward quantum thermal machines for probing thermodynamics in the quantum regime.\\
As an example, in the tapered trap, radial modes for neighboring ions have different frequencies, allowing local addressing in the frequency domain. This enables new ways of implementing quantum simulation and gate operations, such as selectively addressable single-qubit operations in frequency space and exploiting the radial-axial coupling for performing selective two-qubit operations, extending previously utilized quantum bus methods\,\cite{Cirac_Zoller_1995,PhysRevLett.97.050505}.\\
The mechanical nonlinearity opens up new possibilities to advance quantum sensing and metrology through a range of effects, such as bistability, stochastic and vibrational resonances\,\cite{landa2000vibrational}. For instance, coherent signal amplification was achieved by using bistable nanomechanical oscillators\,\cite{Badzey2005}, where the use of noise has led to significant signal amplification. Similarly, we amplified a signal at zeptonewton-level by exploiting vibrational resonances\,\cite{Deng2023}.
Additionally, combining quantum resources such as squeezing with nonlinearities could further increase the sensitivity, similar to effects predicted with quantum thermal machines based on trapped ions\,\cite{Rossnagel2014}.\\
Furthermore, thermodynamics in the quantum regime is a new active research field\,\cite{binder2018thermodynamics,dawkins2018single}, where tapered Paul traps have shown pioneering results with the first single ion heat engine being realized\,\cite{Rossnagel2016} and parameters for work extraction at the Curzon-Ahlborn efficiency limit have been predicted\,\cite{Abah2012}. Quantum resources such as squeezing are expected to increase the efficiency of heat extraction\,\cite{Rossnagel2014} and transient non-confining potentials can be used for speeding up a single ion heat pump through the use of shortcuts to adiabaticity\,\cite{Torrontegui_2018}. The single-ion heat engine can also be used as a sensitive thermal probe and the sensitivity can be enhanced through the application of squeezing\,\cite{Levy_2020}. The tapered trap design also potentially allows for implementing local non-classical baths combining dissipative state preparation\,\cite{Kienzler2015} with frequency space addressing of the local radial modes, which is of fundamental interest in quantum resource theory\,\cite{Lostaglio_2019}.\\

\section*{Acknowledgement}
We thank Florian Elsen and David Zionski for early-stage work on the ion trap setup and simulation. This work was supported by the Deutsche Forschungsgemeinschaft (DFG, German Research Foundation) - Projects No. 499241080, No. 384846402, No. 328961117, No 510794108 - through the QuantERA grant ExTRaQT, the Research Unit Thermal Machines in the Quantum World (FOR 2724), and the Collaborative Research Center ELCH (SFB 1319), and by the federal state of Hesse, Germany, through project SMolBits within the LOEWE program. M.M. and D.W. acknowledge support from Germany’s Excellence Strategy - Cluster of Excellence Matter and Light for Quantum Computing (ML4Q) EXC 2004/1-390534769.\\

\bibliographystyle{iopart-num}
\section*{References}
\bibliography{TechnicalPaperHeatengine}
\end{document}